\documentclass[aps,prl,superscriptaddress,twocolumn,amsmath,amssymb]{revtex4}

\usepackage{color}
\usepackage{graphicx}
\usepackage{bm}
\definecolor{Blue}{rgb}{0.00, 0.00, 1.00}
\definecolor{Red}{rgb}{1.00, 0.00, 0.00}

\begin{document}

\title{Tunable optical properties of multilayers black phosphorus thin films}

\author{Tony Low }
\email{tonyaslow@gmail.com}
\affiliation{IBM T.J. Watson Research Center, 1101 Kitchawan Rd., Yorktown Heights, NY 10598, USA}
\affiliation{Department of Electrical Engineering, Yale University, New Haven, Connecticut 06511 }
\author{ A. S. Rodin}
\affiliation{Boston University, 590 Commonwealth Ave., Boston, MA 02215, USA}
\author{ A. Carvalho }
\affiliation{Graphene Research Centre \& Department of Physics, National University of Singapore, 117542, Singapore}
\author{ Yongjin Jiang }
\affiliation{Zhejiang Normal University, Jinhua 321004, People’s Republic of China}
\author{ Han Wang }
\affiliation{IBM T.J. Watson Research Center, 1101 Kitchawan Rd., Yorktown Heights, NY 10598, USA}
\author{ Fengnian Xia }
\affiliation{Department of Electrical Engineering, Yale University, New Haven, Connecticut 06511 }
\author{ A. H. Castro Neto}
\affiliation{Boston University, 590 Commonwealth Ave., Boston, MA 02215, USA}
\affiliation{Graphene Research Centre \& Department of Physics, National University of Singapore, 117542, Singapore}

\date{\today}
\begin{abstract}
\textcolor[rgb]{0,0,0}{Black phosphorus thin films might offer attractive alternatives to narrow gap compound semiconductors for optoelectronics across mid- to near-infrared frequencies. In this work, we calculate the optical conductivity tensor of multilayer black phosphorus thin films using the Kubo formula within an effective low-energy Hamiltonian. }The optical absorption spectra of multilayer black phosphorus are shown to vary sensitively with thickness, doping, and light polarization. In conjunction with experimental spectra obtained from infrared absorption spectroscopy, we  also discuss the role of interband coupling and disorder on the observed anisotropic absorption spectra. 
\end{abstract}
\maketitle

\section{I. Introduction} The group V element phosphorus can exist in several allotropes, and one of its thermodynamically more stable phases under room temperature and pressure conditions is black phosphorus (BP). Similar to graphene, it is also a layered material, except that each layer forms a puckered surface due to $sp^3$ hybridization. The electrical, optical, and structural properties of single crystalline and polycrystalline BP had been extensively studied in the past\cite{Keyes53,Warschauer63,Jamieson63,morita86review,chang86}. Bulk BP is a semiconductor with a direct band gap of about $0.3\,$eV. Measured Hall mobilities in $n$ and $p-$type samples almost approach $10^5\,$cm$^2$/Vs. In addition, its electrical and optical properties are also highly sensitive to crystallographic orientation. Very recently, BP was re-introduced\cite{Li14BP,Liu14BP,xia14bp,Koenig14} in their multilayer thin film form, obtained from the simple mechanical exfoliation\cite{novoselov2d05}. Preliminary electrical data on multilayer BP field-effect transistors showed encouraging results.

In this work, we examine the optical properties of \textcolor[rgb]{0,0,0}{multilayer BP thin films} with thickness ranging from few to tens of nanometers. We calculate its optical conductivities using the Kubo formula, within the framework of an effective low-energy Hamiltonian\cite{rodin14}. Our calculations show that the optical absorption spectra of multilayer BP vary sensitively with thickness, doping, and light polarization, especially across the technologically relevant mid- to near-infrared spectrum. In conjunction with experimental spectra obtained from Fourier transform infrared spectroscopy (FTIR)\cite{xia14bp}, we elucidate the role of interband coupling and disorder on the observed anisotropic absorption spectra.

\section{II. Model} 
\begin{figure}[t]
\centering
\scalebox{0.55}[0.55]{\includegraphics*[viewport=170 185 800 460]{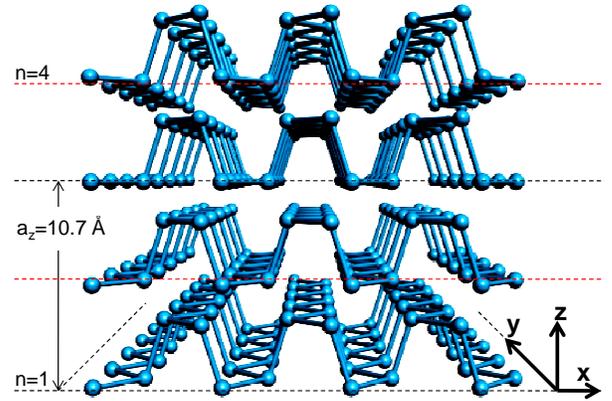}}
\caption{ Lattice structure of black phosphorus(BP). Layer numbers $n$ is indicated. $a_z$ is the lattice constant in the out-of-plane direction. Thickness of the multilayer BP is then given by $t_z=n*a_z/2$. Crystallographic axes $\{ xyz \}$ used in this work are defined. 
}
\label{fig1}
\end{figure}

BP has an orthorhombic crystal structure consisting of puckered layers as illustrated in Fig.\,\ref{fig1}. Lattice constant in the out-of-plane direction is about $10.7\,$\AA, where effective layer-to-layer distance is half of this. In multilayer BP, translational symmetry in the $z$-direction is broken, and its bandstructure has a direct energy gap at the $\Gamma$ point instead of the Z point in bulk.  In order to describe the system behavior around the $\Gamma$ point, we use first principles calculations in conjunction with the $\mathbf{k}$$\cdot$$\mathbf{p}$ approximation\cite{rodin14}. 

Before constructing the low-energy Hamiltonian, we analyze the system symmetry. Individual monolayers, due to puckering, have a reduced symmetry compared to graphene, represented by the $C_{2h}$ point group. The principal axis, which we denote $\hat y$, runs along the buckles. The reflection plane $\sigma_h$ lies in the $x$-$z$ plane. It is possible to show that at the $\Gamma$ point wavefunctions are either even or odd with respect to reflection across $\sigma_h$. From \emph{ab initio} calculations, we determined that the highest valence and the lowest conduction bands are even and composed primarily of $s$, $p_x$, and $p_z$ orbitals with a small contribution from even $d$ orbitals. Odd wavefunctions, made up of mostly $p_y$, are more energetically separated from the Fermi level. Keeping this in mind, we now proceed to construct an effective Hamiltonian that describes the coupled valence and conduction bands.

In $\mathbf{k}$$\cdot$$\mathbf{p}$ approximation, the perturbing Hamiltonian is given by $\mathcal{H}_1 =\hbar(k_x\hat p_x+k_y\hat p_y)/m_0$. To determine the coupling between the bands, one needs to compute the matrix elements of $\mathcal{H}_1$ for the two bands of interest. Since $\hat p_x$ is even with respect to $\sigma_h$ reflection, it gives a finite contribution to the matrix element. On the other hand, the fact that $\hat p_y$ is odd prohibits the term linear in $k_y$. If one disregards the remaining bands, the system becomes quasi-one-dimensional close to the $\Gamma$ point. To introduce $k_y$ dependence, we utilize the L\"{o}wdin partitioning. The leading order correction to the effective Hamiltonian is given by
\begin{equation}
(\mathcal{H}_1^{(2)})_{mm'} = \sum_l\frac{\left(\mathcal H_1\right)_{ml}\left(\mathcal H_1\right)_{lm'}}{2}\left[\frac{1}{E_m-E_l}+\frac{1}{E_{m'}-E_l}\right]\,,
\end{equation}
where the summation goes over the remaining bands. Because of the product of the matrix elements in the numerator, it is clear that the leading order $k_y$ contribution to the coupling terms will be quadratic, arising from the mixing with the odd bands. This allows one to write down the low-energy in-plane Hamiltonian around the $\Gamma$ point as\cite{rodin14},
\begin{eqnarray}
{\cal H} = \left(
\begin{array}{cc}
E_c + \eta_c k_x^2 + \nu_c k_y^2 & \gamma k_x + \beta k_y^2\\
\gamma k_x + \beta k_y^2 & E_v - \eta_v k_x^2 - \nu_v k_y^2
\end{array} \right)
\label{hamil}
\end{eqnarray}
where $\eta_{c,v}$ and $\nu_{c,v}$ are related to the effective masses, while $\gamma$ and $\beta$ describe the effective couplings between the conduction and valence bands. $E_c$ and $E_v$ are the conduction and valence band edge energies in bulk BP, where $E_c-E_v$ is the bulk gap of $\approx 0.3\,$eV\cite{morita86review}.

\textcolor[rgb]{0,0,0}{Electrons in BP are energetically highly dispersive and delocalized along the out-of-plane direction \cite{han14arpes,Takahashi86arpes}, unlike other layered materials such as graphene and the transition metal dichalcogenides (TMDs). Angle-resolved photoemission spectroscopy (ARPES) studies confirm these parabolic-like out-of-plane $\Gamma-Z$ dispersion in BP, with an energy bandwidth of $\sim 0.7\,$eV for the valence band\cite{han14arpes}. Cyclotron resonance experiments on bulk BP\cite{Narita83cy} also found the out-of-plane effective masses to be considerably smaller than that of TMDs\cite{Mattheiss73} and graphite\cite{wallace47}. In this work, we adopt an average of experimental\cite{Narita83cy} and theoretically\cite{Narita83cy} (See accompanying Suppl. Info) predicted electron and hole out-of-plane mass i.e. $m_{cz}\approx 0.2\,m_0$ and $m_{vz}\approx 0.4\,m_0$. For multilayer BP thin films, confinement in the out-of-plane $z$ direction leads to multiple subbands, $E_{c,v}^j$. The in-plane dispersion within each subband $j$ can be described by Eq.\,\eqref{hamil}, where $E_{c,v}$ are being replaced with $E_{c,v}^j$. More explicitly, the additional confinement energies, $\delta E_{c}^j$, are given by $j^2\hbar^2\pi^2/2m_{cz}t_z^2+\delta_c(t_z)$, where $j$ labels the subband, $t_z$ is the thickness of the BP film, and $m_{cz}$ is the electron effective mass along $z$. Analogous expressions apply also for the hole case. The quantities $\delta_{c,v}(t_z)$ are chosen such that they reproduce the predicted energy gap of the BP film\cite{Tran14}, of 2\,eV and 0.3\,eV in the monolayer and bulk limit respectively. In this work, we restrict ourselves to BP thin films $4\,$nm and larger, where the confinement energies are within the energy bandwidth of $\sim 0.7\,$eV where the effective mass approximation is valid.   }

\textcolor[rgb]{0,0,0}{Close to the $\Gamma$ point, the band parameters are related to the in-plane effective masses via\cite{rodin14},
}\begin{eqnarray}
m_{cx}^{j}=\frac{\hbar^2}{2\gamma^2/(E_c^j-E_v^j)+\eta_{c}} &\mbox{   ,   }& m_{cy}^j=\frac{\hbar^2}{2\nu_c}
\end{eqnarray}
The band parameters $\eta_{c,v}$, $\nu_{c,v}$ and $\gamma$ are chosen such that they yield the known effective masses in the bulk BP limit i.e. $m_{cx}=m_{vx}=0.08\,m_0$, $m_{cy}=0.7\,m_0$ and $m_{vy}=1.0\,m_0$\cite{morita86review,Narita83cy}, and $m_{cx}=m_{vx}\approx 0.15\,m_0$ for monolayer BP\cite{rodin14} (with an estimated bandgap of $\sim 2\,$eV). \textcolor[rgb]{0,0,0}{The value of $\beta$ was suggested to lie between $1-10a^2/\pi^2\,$eVm$^2$\cite{rodin14}, where $a\approx 2.23$ \AA\, and $\pi/a$ is the width of the BZ in $x$ direction. }We tentatively assign $\beta\approx 2a^2/\pi^2\,$eVm$^2$ and evaluate its effect on our calculated results in subsequent discussions below.

Physical quantities observed in optical experiments can often be expressed in terms of the optical conductivity. The Kubo formula for the conductivity tensor as function of frequency and momentum reads,
\begin{eqnarray}
\nonumber
\sigma_{\alpha\beta}(\textbf{q},\omega)&&=-i\frac{g_s\hbar e^2}{(2\pi)^2}\sum_{ss'jj'} \int d\textbf{k} \frac{f(E_{sj\textbf{k}})-f(E_{s'j'\textbf{k}'})}{E_{sj\textbf{k}}-E_{s'j'\textbf{k}'}}\\
&&\times\frac{\left\langle \Phi_{sj\textbf{k}}\right|\hat{v}_{\alpha}\left|\Phi_{s'j'\textbf{k}'}\right\rangle \left\langle \Phi_{s'j'\textbf{k}'}\right|\hat{v}_{\beta}\left|\Phi_{sj\textbf{k}}\right\rangle}{E_{sj\textbf{k}}-E_{s'j'\textbf{k}'}+\hbar\omega+i\eta}
\label{kubofor}
\end{eqnarray}
where $\hat{v}_{\alpha}$ is the velocity operator defined as $\hbar^{-1}\partial_{k\alpha}{\cal H}$, $g_s=2$ accounts for the spin degeneracy and $\eta\approx 10\,$meV accounts for the finite damping. $f(...)$ is the Fermi-Dirac distribution function, where temperature is taken to be $300\,$K in all calculations. The indices $\{s,s'\}=\pm 1$ denote conduction/valence band, while $\{j,j'\}$ are the subbands indices. $E_{sj\textbf{k}}$ and $\Phi_{sj\textbf{k}}$ are the eigen-energies and eigen-functions of ${\cal H}$. We have an analytical expression for $E_{sj\textbf{k}}$; 
\begin{eqnarray}
\nonumber
E_{\pm j\textbf{k}} = \tfrac{1}{2}\left[(E_c^j+E_v^j)+k_x^2(\eta_c-\eta_v)+k_y^2(\nu_c-\nu_v)\right]\\
\nonumber
\pm\tfrac{1}{2}\left[\Delta^2+\Delta\left(2k_x^2(\eta_c+\eta_v)+2k_y^2(\nu_c+\nu_v)\right)\right.\\
\left.+\left(k_x^2(\eta_c+\eta_v)+k_y^2(\nu_c+\nu_v)\right)^2+4(\gamma k_x+\beta k_y^2)^2\right]^{1/2}
\end{eqnarray}
where $\Delta\equiv E_c^j-E_v^j$. Optical transitions between these quantized subbands are allowed when $ss'=\pm 1$ (i.e. intra- and inter-band processes) and $j=j'$. Otherwise the matrix elements $\left\langle ...\right\rangle$ in Eq.\,\ref{kubofor} vanish. In this work, we are only interested in the local conductivity i.e. $\sigma_{\alpha\beta}(\textbf{q}\rightarrow 0,\omega)$. In this limit, only the diagonal components of the conductivity tensor, $\sigma_{xx}(\omega)$ and $\sigma_{yy}(\omega)$, are non-zero.

\begin{figure}[t]
\centering
\scalebox{0.75}[0.75]{\includegraphics*[viewport=220 110 800 540]{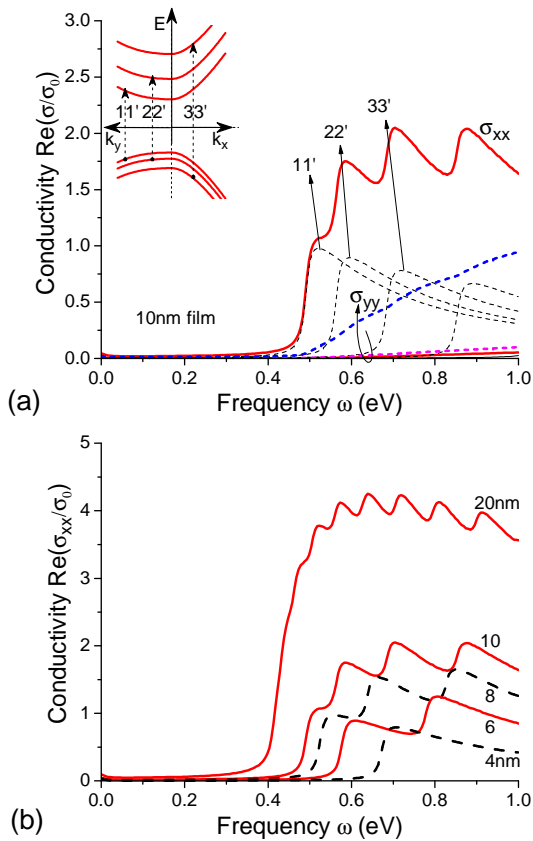}}
\caption{ \textbf{(a)} Real part of optical conductivities, Re($\sigma_{xx}$) and Re($\sigma_{yy}$), for $10\,$nm thick intrinsic BP, i.e. Fermi level is located at mid-gap. See main text for parameters used in the calculations. Inter-subbands contributions to the $\sigma_{xx}$ are plotted, where the respective optical transition processes are illustrated in inset. $\sigma_{yy}$ calculated with $2\times$ and $10\times$ the value of $\beta$ are also shown (dashed pink and blue lines respectively) for comparison. Conductivities are normalized with respect to $\sigma_0\equiv e^2/4\hbar$.
\textbf{(b)} Re($\sigma_{xx}$) for intrinsic BP of different thicknesses as indicated.
}
\label{fig2}
\end{figure}

\begin{figure}[htps]
\centering
\scalebox{0.75}[0.75]{\includegraphics*[viewport=240 190 510 510]{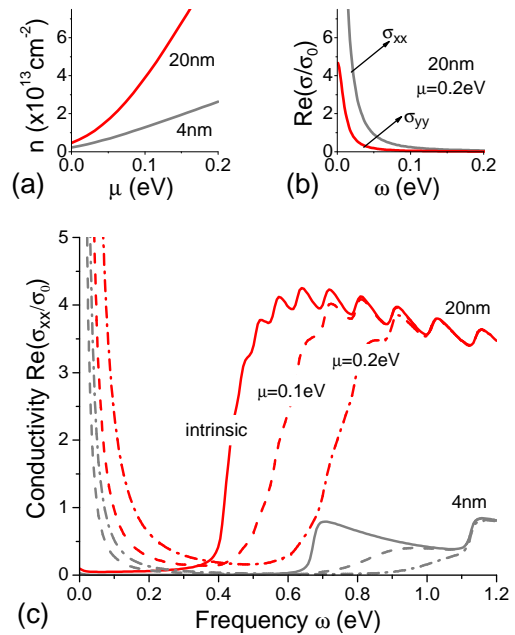}}
\caption{  \textbf{(a)} Electron densities ($n$) as function of chemical potential ($\mu$) for $20\,$nm and $4\,$nm BP films. See text for definition of $\mu$. \textbf{(b)} Optical conductivity $\sigma_{xx}$ and $\sigma_{yy}$ for $20\,$nm BP film doped at $\mu=0.2\,$eV. \textbf{(c)} Re($\sigma_{xx}$) for BP with different chemical potential $\mu$ as indicated, calculated for $4\,$nm and $20\,$nm thick films. 
}
\label{fig3}
\end{figure}

\section{III. Results} 
\subsection{A. Thickness dependence} 
Fig.\,\ref{fig2}a presents the calculated real part of optical conductivities of an undoped $10\,$nm BP thin film. Temperature is taken to be $300\,$K in all calculations. Results are normalized with respect to $\sigma_0=e^2/4\hbar$, the well-known universal conductivity of graphene\cite{nair08,mak08mea}. The large asymmetry between $\sigma_{xx}$ and $\sigma_{yy}$ is immediately apparent. We note that this asymmetry has less to do with the in-plane effective mass anisotropy, as the qualitative trends persist even if we set $m_x=m_y$ (not shown). Instead, $\sigma_{yy}$ increases almost linearly with the interband coupling term, $\beta$, as shown (pink and blue dashed lines). However, within the reasonable range of $\beta$ values from $1-10a^2/\pi^2\,$eVm$^2$, the magnitude of $\sigma_{xx}$ remains relatively unchanged.

As shown in Fig.\,\ref{fig2}a, $\sigma_{xx}$ exhibits an oscillatory behavior with $\omega$ which can be traced to the underlying electronic subbands structure. Since it is undoped, only interband transitions from \textcolor[rgb]{0,0,0}{the valence to conduction} band are permitted. The zero wavefunction overlap between different subbands implies that only $j=j'$ transitions are allowed, as illustrated in the inset. \textcolor[rgb]{0,0,0}{The structure in the observed subband spectra is related to the joint density-of-states, which resembles that of a one-dimensional system like carbon nanotubes, and is a direct consequence of the large band anisotropy in BP. }

\begin{figure*}[t]
\centering
\scalebox{0.72}[0.72]{\includegraphics*[viewport=55 205 810 400]{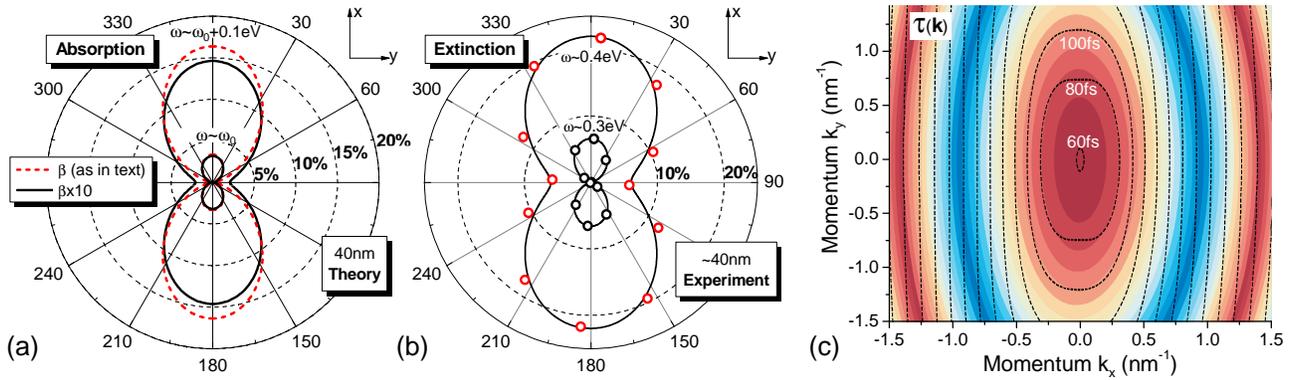}}
\caption{   \textbf{(a)} Polar representation of the absorption coefficient ${\cal A}(\alpha)$ for a \textcolor[rgb]{0,0,0}{$40\,$nm} intrinsic BP film for normal incident light with excitation energies at the band gap $\omega_0$, and larger. $\alpha$ is the light polarization angle. ${\cal A}(\alpha)$ is plotted for two values of interband coupling strengths. \textbf{(b)} Polar representation of the experimental extinction spectra ${\cal Z}(\omega)$ obtained from FTIR spectroscopy, for a $\sim 40\,$nm BP film on a SiO$_2$ substrate\cite{xia14bp}. We estimate that the beam spot size is $\approx 2$ times larger than the sample size in experiment. Lines are fitted curves to the data using $a\mbox{sin}^2\theta+b\mbox{cos}^2\theta$. \textbf{(c)} Contour plots for single particle lifetime due to long-range Coulomb interaction, overlaid with the energy dispersion colour-filled contours. Each color step has an energy difference of $20\,$meV. See accompanying text for more descriptions on (b) and (c).
}
\label{fig4}
\end{figure*}

Fig.\,\ref{fig2}b shows the calculated $\sigma_{xx}$ for BP films of different thicknesses $t_z$ from $4-20\,$nm. The absorption edge moves from $0.3\,$eV to $0.6\,$eV with decreasing $t_z$ due to the increasing energy gap. Maximal $\sigma_{xx}$ also decreases as a result. \textcolor[rgb]{0,0,0}{At frequencies larger than the absorption edge, the optical conductivity  roughly saturates to some value for the $20\,$nm film. In the freestanding case, this translates to an absorption of $\sim 8\,\%$, which increases linearly with thickness at a rate of $0.4\,\%$ per nm for thicker films. } Due to reduced screening for \textcolor[rgb]{0,0,0}{films thinner than those studied in this paper}, strong excitonic effects is expected\cite{qiu13gw}, which would lead to enhanced light absorption \textcolor[rgb]{0,0,0}{near the absorption edge} not accounted for in present calculations.

\subsection{B. Doping dependence}

Doping can be induced either electrically or chemically by introducing donor or acceptor impurity atoms during the synthesis\cite{morita86review}. Here, we consider the uniformly $n$-doped case. We defined the chemical potential $\mu$ to be the difference between the Fermi level and the first conduction subband i.e. $E_F-E_c^1$. The electron density, $n$, follows from Fermi statistics and is given by,
\begin{eqnarray}
n=\sum_j \frac{m_{dos}^j k_B T}{\pi\hbar^2}\mbox{log}\left[1+\mbox{exp}\left(\frac{E_F-E_c^j}{k_B T}\right)\right]
\end{eqnarray}
where $m_{dos}^j=(m_{cx}^j m_{cy}^j)^{1/2}$ is the density-of-states mass. Fig.\,\ref{fig3}a shows $n$ as a function of $\mu$ for $4\,$nm and $20\,$nm films, where $\mu\lesssim 0.2\,$eV covers the range of dopings $\lesssim 5\times 10^{13}\,$cm$^{-2}$. These dopings are routinely obtained in experiments with layered materials. Following the increase in doping will be the appearance of Drude absorption peak at $\omega\rightarrow 0$. Fig.\,\ref{fig3}b plots the optical conductivities, $\sigma_{xx}$ and $\sigma_{yy}$, for a doped $20\,$nm BP film. The asymmetry here is due to the larger Drude weight ${\cal D}$ for transport along $x$, to be discussed below.

Fig.\,\ref{fig3}c studies the optical property of multilayer BP with doping ranging from $\mu=0-0.2\,$eV. Simple picture for direct optical transitions would suggest a blue-shift in the absorption edge by $2\mu$ due to Pauli blocking. Consider first the result for $20\,$nm film. Although the blue-shift is evident, it appears to be less than $2\mu$ \textcolor[rgb]{0,0,0}{and more smeared out in energy due to finite temperature effects}. The absorption edge also becomes less abrupt with increased doping. These trends become more pronounced in the $4\,$nm case. 

\subsection{C. Angular dependence} 

Light scattering across a conducting layer between two dielectric media can be solved via the Maxwell equation. Due to the anisotropic conductivity of BP, unless the incident light is polarized along one of the crystal axes, the reflected polarization will be different from the incident one. However, when the surface conductivity is much smaller than the free space admittance as in our case, i.e. $\sigma_{\alpha\alpha}\ll \sqrt{\epsilon_0/\mu_0}$, this polarization rotation can be neglected. Therefore, for normal incidence, the reflectivity for the unrotated component is given by
\begin{eqnarray}
r=-\frac{\epsilon_0 c(\sqrt{\epsilon_2}-\sqrt{\epsilon_1})+\sigma_{xx} \mbox{cos}^2\alpha +\sigma_{yy}\mbox{sin}^2\alpha}{\epsilon_0 c(\sqrt{\epsilon_2}+\sqrt{\epsilon_1})+\sigma_{xx} \mbox{cos}^2\alpha +\sigma_{yy}\mbox{sin}^2\alpha}
\end{eqnarray}
where $\epsilon_0$ is the free-space permittivity, $c$ the speed of light and $\alpha$ is the light polarization angle. The reflection and transmission probabilities are given by ${\cal R}\approx|r|^2$ and ${\cal T}\approx |1+r|^2 \sqrt{\epsilon_2/\epsilon_1}$, and the absorption coefficient is ${\cal A}=1-{\cal R}-{\cal T}$.

Fig.\,\ref{fig4}a plots the angle-dependent absorption coefficient, ${\cal A}(\alpha)$, for an intrinsic $\textcolor[rgb]{0,0,0}{40}\,$nm BP film, with light excitations equal and larger than the bandgap i.e. interband processes. We observed that ${\cal A}(\alpha)$ exhibits strong dependence on polarization angle. The absorption anisotropy is rather sensitive to the interband coupling strength $\beta$, which evolves from a `dumbbell'-shape to a more elliptical form with increasing $\beta$. Disorder, as described by the phenomenological constant $\eta$ in Eq.\,\ref{kubofor}, has a milder effect (not shown). 

FTIR spectroscopy is a common technique for obtaining infrared spectra of absorption. In Fig.\,\ref{fig4}b, we show the experimental extinction spectra ${\cal Z}(\omega)$ for a $40\,$nm BP film on a SiO$_2$ substrate. The extinction characterizes the differential transmission through the sample in regions with (${\cal T}(\omega)$) and without (${\cal T}_0(\omega)$) BP i.e. ${\cal Z}=1-{\cal T}/{\cal T}_0$. Experimental setup and details are elaborated elsewhere\cite{xia14bp}. Note that ${\cal Z}(\omega)$ includes both the additional light absorption and reflection due to the BP, ${\cal Z}(\omega)>{\cal A}(\omega)$. To account for the observed anisotropy in the experimental spectra, the model would entail an interband coupling strength of $\beta \approx 20a^2/\pi^2\,$eVm$^2$ as depicted in Fig.\,\ref{fig4}a. This allows us to capture the main experimental features observed. It is very likely that extrinsic factors such as structural deformations, corrugations or defects also play a role in the interband optical conductivity anisotropy, which we defer to future studies.

\subsection{D. Disorder} 

Contrary to the interband case, intraband optical absorption in doped samples are primarily dictated by disorder. Due to the mass anisotropy, the Drude weight is different along the two crystallographic axes, and is given by ${\cal D}_{j}=\pi n e^2/m_{j}$ where $j=\{x,y\}$. In the zero and large frequencies limits, the real part of Drude conductivity are given by $\hbar{\cal D}_{j}/\pi\eta$ and ${\cal D}_j\eta/\pi\hbar\omega^2$ respectively. Below, we discuss some common sources of disorders on the intraband conductivity anisotropy.

Impurities and defects are common sources of disorder in 2D materials\cite{neto09rmp,sarma11rmp}. For interactions of electrons with impurities, the self-energy within the Born approximation can be written as, 
\begin{eqnarray}
\Sigma(\textbf{k},E)=n_i\frac{g_s}{(2\pi)^2}\int d\textbf{k} \frac{|V(q)|^2{\cal F}(\textbf{k},\textbf{k}')}{E-E_{\textbf{k}'}+i\eta_0}
\end{eqnarray}
where $V(q)$ is the scattering potential, ${\cal F}(\textbf{k},\textbf{k}')\equiv |\left\langle\textbf{k}\right.\left|\textbf{k}'\right\rangle|^2$ is the overlap function between states and $\eta_0$ is taken to be $10\,$meV. For long-range Coulomb interaction, we have $V(q)=e^2 \mbox{exp}(-qd_0)/2\epsilon_0\kappa(q+q_s)$, where $\kappa$ is the effective dielectric constant taken to be $1$ and $d_0$ is an effective distance between BP and the impurities. \textcolor[rgb]{0,0,0}{Following Ref. \cite{low14bpplasmon}, screening is relatively isotropic for the relevant $q$ considered here, and can be treated within the Thomas-Fermi approximation with $q_s=e^2 D/2\pi\kappa$ being the screening wave-vector, and $D$ is the 2D density-of-states.} For short-range interaction, we simply have $V(q)=V_0$. Unlike the latter, long-range Coulomb interaction would lead to single particle damping that is angular dependent. 

The single particle lifetime is given by $\tau(\textbf{k})=\tfrac{\hbar}{2}\mbox{Im}[\Sigma(\textbf{k},E)^{-1}]$, a quantity that can be directly measured with ARPES. In Fig.\,\ref{fig4}c, we plot the contours for single particle lifetime due to long-range Coulomb interaction. For our estimates, we assume typical impurities concentration of $n_i\approx 5\times 10^{11}\,$cm$^{-2}$ and $d_0\approx 1\,$nm. The calculated lifetime contours, $\tau(\textbf{k})$, is overlaid with the energy dispersion contours, and they are qualitatively different. This implies that electron interaction with long-range Coulomb impurities will lead to an anisotropic lifetime in general, and subsequently influencing also the conductivity anisotropy. In similar fashion, interaction with acoustic phonons would also yield an anisotropic single particle lifetime due to the different sound velocities and deformation potentials along the two crystal axes.

\section{IV. Conclusions} 
In summary, we have presented a model for describing the optical properties in multilayer BP. \textcolor[rgb]{0,0,0}{In particular, we focus primarily on films where its effective bandgap resides in mid- to near-infrared frequencies, with thicknesses of $\sim 10-60$ layers. The simple model proposed here facilitates the modeling of experimental optical spectra of BP thin films, which can include also the effect of disorders that will be hardly tractable by \emph{ab initio} methods.} Our results show that the optical conductivity, similarly the absorption spectra, of multilayer BP vary sensitively with thickness, doping and light polarization, particularly for frequencies ranging from $2500-5000\,$cm$^{-1}$, which resides in the technologically relevant mid- to near-infrared spectrum. Hence, multilayer BP might offer attractive alternatives, in terms of tunability, flexibility and cost, to narrow gap compound semiconductors for infrared optoelectronics.  

%

\emph{Acknowledgements---} 
T.L. is grateful for the hospitality of Graphene Research Center, Singapore, where this work was initiated and conducted in large part. A.S.R. acknowledges DOE grant DE-FG02-08ER46512, ONR grant MURI N00014-09-1-1063. A.H.C.N. acknowledges NRF-CRP award ``Novel 2D materials with tailored properties: beyond graphene'' (R-144-000-295-281). The first-principles calculations were carried out on the GRC high-performance computing facilities.

\emph{Supplemental Information---} Details of \emph{ab initio} studies with extracted mass tensors are summarized.


\begin{thebibliography}{23}
\expandafter\ifx\csname natexlab\endcsname\relax\def\natexlab#1{#1}\fi
\expandafter\ifx\csname bibnamefont\endcsname\relax
  \def\bibnamefont#1{#1}\fi
\expandafter\ifx\csname bibfnamefont\endcsname\relax
  \def\bibfnamefont#1{#1}\fi
\expandafter\ifx\csname citenamefont\endcsname\relax
  \def\citenamefont#1{#1}\fi
\expandafter\ifx\csname url\endcsname\relax
  \def\url#1{\texttt{#1}}\fi
\expandafter\ifx\csname urlprefix\endcsname\relax\def\urlprefix{URL }\fi
\providecommand{\bibinfo}[2]{#2}
\providecommand{\eprint}[2][]{\url{#2}}

\bibitem[{\citenamefont{Keyes}(1953)}]{Keyes53}
\bibinfo{author}{\bibfnamefont{R.~W.} \bibnamefont{Keyes}},
  \bibinfo{journal}{Physical Review} \textbf{\bibinfo{volume}{92}},
  \bibinfo{pages}{580} (\bibinfo{year}{1953}).

\bibitem[{\citenamefont{Warschauer}(1963)}]{Warschauer63}
\bibinfo{author}{\bibfnamefont{D.}~\bibnamefont{Warschauer}},
  \bibinfo{journal}{Journal of Applied Physics} \textbf{\bibinfo{volume}{34}},
  \bibinfo{pages}{1853} (\bibinfo{year}{1963}).

\bibitem[{\citenamefont{Jamieson}(1963)}]{Jamieson63}
\bibinfo{author}{\bibfnamefont{J.~C.} \bibnamefont{Jamieson}},
  \bibinfo{journal}{Science} \textbf{\bibinfo{volume}{139}},
  \bibinfo{pages}{1291} (\bibinfo{year}{1963}).

\bibitem[{\citenamefont{Morita}(1986)}]{morita86review}
\bibinfo{author}{\bibfnamefont{A.}~\bibnamefont{Morita}},
  \bibinfo{journal}{Applied Physics A} \textbf{\bibinfo{volume}{39}},
  \bibinfo{pages}{227} (\bibinfo{year}{1986}).

\bibitem[{\citenamefont{Chang and Cohen}(1986)}]{chang86}
\bibinfo{author}{\bibfnamefont{K.~J.} \bibnamefont{Chang}} \bibnamefont{and}
  \bibinfo{author}{\bibfnamefont{M.~L.} \bibnamefont{Cohen}},
  \bibinfo{journal}{Physical Review B: Condensed Matter}
  \textbf{\bibinfo{volume}{33}}, \bibinfo{pages}{6177} (\bibinfo{year}{1986}).

\bibitem[{\citenamefont{Li et~al.}(2014)\citenamefont{Li, Yu, Ye, Ge, Ou, Wu,
  Feng, Chen, and Zhang}}]{Li14BP}
\bibinfo{author}{\bibfnamefont{L.}~\bibnamefont{Li}},
  \bibinfo{author}{\bibfnamefont{Y.}~\bibnamefont{Yu}},
  \bibinfo{author}{\bibfnamefont{G.~J.} \bibnamefont{Ye}},
  \bibinfo{author}{\bibfnamefont{Q.}~\bibnamefont{Ge}},
  \bibinfo{author}{\bibfnamefont{X.}~\bibnamefont{Ou}},
  \bibinfo{author}{\bibfnamefont{H.}~\bibnamefont{Wu}},
  \bibinfo{author}{\bibfnamefont{D.}~\bibnamefont{Feng}},
  \bibinfo{author}{\bibfnamefont{X.~H.} \bibnamefont{Chen}}, \bibnamefont{and}
  \bibinfo{author}{\bibfnamefont{Y.}~\bibnamefont{Zhang}},
  \bibinfo{journal}{Nature Nanotechnology}  (\bibinfo{year}{2014}).

\bibitem[{\citenamefont{Liu et~al.}(2014)\citenamefont{Liu, Neal, Zhu, Tomanek,
  and Ye}}]{Liu14BP}
\bibinfo{author}{\bibfnamefont{H.}~\bibnamefont{Liu}},
  \bibinfo{author}{\bibfnamefont{A.~T.} \bibnamefont{Neal}},
  \bibinfo{author}{\bibfnamefont{Z.}~\bibnamefont{Zhu}},
  \bibinfo{author}{\bibfnamefont{D.}~\bibnamefont{Tomanek}}, \bibnamefont{and}
  \bibinfo{author}{\bibfnamefont{P.~D.} \bibnamefont{Ye}},
  \bibinfo{journal}{arXiv:1401.4133}  (\bibinfo{year}{2014}).

\bibitem[{\citenamefont{Xia et~al.}(2014)\citenamefont{Xia, Wang, and
  Jia}}]{xia14bp}
\bibinfo{author}{\bibfnamefont{F.}~\bibnamefont{Xia}},
  \bibinfo{author}{\bibfnamefont{H.}~\bibnamefont{Wang}}, \bibnamefont{and}
  \bibinfo{author}{\bibfnamefont{Y.}~\bibnamefont{Jia}},
  \bibinfo{journal}{arXiv:1402.0270}  (\bibinfo{year}{2014}).

\bibitem[{\citenamefont{Koenig et~al.}(2014)\citenamefont{Koenig, Doganov,
  Schmidt, Neto, and Oezyilmaz}}]{Koenig14}
\bibinfo{author}{\bibfnamefont{S.~P.} \bibnamefont{Koenig}},
  \bibinfo{author}{\bibfnamefont{R.~A.} \bibnamefont{Doganov}},
  \bibinfo{author}{\bibfnamefont{H.}~\bibnamefont{Schmidt}},
  \bibinfo{author}{\bibfnamefont{A.~H.} \bibnamefont{Neto}}, \bibnamefont{and}
  \bibinfo{author}{\bibfnamefont{B.}~\bibnamefont{Oezyilmaz}},
  \bibinfo{journal}{Appl. Phys. Lett.} \textbf{\bibinfo{volume}{104}},
  \bibinfo{pages}{103106} (\bibinfo{year}{2014}).

\bibitem[{\citenamefont{Novoselov et~al.}(2005)\citenamefont{Novoselov, Jiang,
  Schedin, Booth, Khotkevich, Morozov, and Geim}}]{novoselov2d05}
\bibinfo{author}{\bibfnamefont{K.~S.} \bibnamefont{Novoselov}},
  \bibinfo{author}{\bibfnamefont{D.}~\bibnamefont{Jiang}},
  \bibinfo{author}{\bibfnamefont{F.}~\bibnamefont{Schedin}},
  \bibinfo{author}{\bibfnamefont{T.~J.} \bibnamefont{Booth}},
  \bibinfo{author}{\bibfnamefont{V.~V.} \bibnamefont{Khotkevich}},
  \bibinfo{author}{\bibfnamefont{S.~V.} \bibnamefont{Morozov}},
  \bibnamefont{and} \bibinfo{author}{\bibfnamefont{A.~K.} \bibnamefont{Geim}},
  \bibinfo{journal}{Proceedings of the National Academy of Sciences of the
  United States of America} \textbf{\bibinfo{volume}{102}},
  \bibinfo{pages}{10451} (\bibinfo{year}{2005}).

\bibitem[{\citenamefont{Rodin et~al.}(2014)\citenamefont{Rodin, Carvalho, and
  Neto}}]{rodin14}
\bibinfo{author}{\bibfnamefont{A.~S.} \bibnamefont{Rodin}},
  \bibinfo{author}{\bibfnamefont{A.}~\bibnamefont{Carvalho}}, \bibnamefont{and}
  \bibinfo{author}{\bibfnamefont{A.~H.} \bibnamefont{Neto}},
  \bibinfo{journal}{arXiv:1401.1801}  (\bibinfo{year}{2014}).

\bibitem[{\citenamefont{Han et~al.}(2014)\citenamefont{Han, Yao, Bai, Miao,
  Zhu, Guan, and al.}}]{han14arpes}
\bibinfo{author}{\bibfnamefont{C.~Q.} \bibnamefont{Han}},
  \bibinfo{author}{\bibfnamefont{M.~Y.} \bibnamefont{Yao}},
  \bibinfo{author}{\bibfnamefont{X.~X.} \bibnamefont{Bai}},
  \bibinfo{author}{\bibfnamefont{L.}~\bibnamefont{Miao}},
  \bibinfo{author}{\bibfnamefont{F.}~\bibnamefont{Zhu}},
  \bibinfo{author}{\bibfnamefont{D.~D.} \bibnamefont{Guan}}, \bibnamefont{and}
  \bibinfo{author}{\bibfnamefont{S.~W.~e.} \bibnamefont{al.}},
  \bibinfo{journal}{arXiv preprint arXiv:1405.7431}  (\bibinfo{year}{2014}).

\bibitem[{\citenamefont{Takahashi et~al.}(1986)\citenamefont{Takahashi,
  Gunasekara, Ohsawa, Ishii, Kinoshita, Suzuki, Sagawa, Kato, Miyahara, and
  Shirotani}}]{Takahashi86arpes}
\bibinfo{author}{\bibfnamefont{T.}~\bibnamefont{Takahashi}},
  \bibinfo{author}{\bibfnamefont{N.}~\bibnamefont{Gunasekara}},
  \bibinfo{author}{\bibfnamefont{H.}~\bibnamefont{Ohsawa}},
  \bibinfo{author}{\bibfnamefont{H.}~\bibnamefont{Ishii}},
  \bibinfo{author}{\bibfnamefont{T.}~\bibnamefont{Kinoshita}},
  \bibinfo{author}{\bibfnamefont{S.}~\bibnamefont{Suzuki}},
  \bibinfo{author}{\bibfnamefont{T.}~\bibnamefont{Sagawa}},
  \bibinfo{author}{\bibfnamefont{H.}~\bibnamefont{Kato}},
  \bibinfo{author}{\bibfnamefont{T.}~\bibnamefont{Miyahara}}, \bibnamefont{and}
  \bibinfo{author}{\bibfnamefont{I.}~\bibnamefont{Shirotani}},
  \bibinfo{journal}{Physical Review B} \textbf{\bibinfo{volume}{33}},
  \bibinfo{pages}{4324} (\bibinfo{year}{1986}).

\bibitem[{\citenamefont{Narita et~al.}(1983)\citenamefont{Narita, Terada, Mori,
  Muro, Akahama, and Endo.}}]{Narita83cy}
\bibinfo{author}{\bibfnamefont{S.-i.} \bibnamefont{Narita}},
  \bibinfo{author}{\bibfnamefont{S.-i.} \bibnamefont{Terada}},
  \bibinfo{author}{\bibfnamefont{S.}~\bibnamefont{Mori}},
  \bibinfo{author}{\bibfnamefont{K.}~\bibnamefont{Muro}},
  \bibinfo{author}{\bibfnamefont{Y.}~\bibnamefont{Akahama}}, \bibnamefont{and}
  \bibinfo{author}{\bibfnamefont{S.}~\bibnamefont{Endo.}},
  \bibinfo{journal}{Journal of the Physical Society of Japan}
  \textbf{\bibinfo{volume}{52}}, \bibinfo{pages}{3544} (\bibinfo{year}{1983}).

\bibitem[{\citenamefont{Mattheiss}(1973)}]{Mattheiss73}
\bibinfo{author}{\bibfnamefont{L.~F.} \bibnamefont{Mattheiss}},
  \bibinfo{journal}{Physical Review B} \textbf{\bibinfo{volume}{8}},
  \bibinfo{pages}{3719} (\bibinfo{year}{1973}).

\bibitem[{\citenamefont{Wallace}(1947)}]{wallace47}
\bibinfo{author}{\bibfnamefont{P.~R.} \bibnamefont{Wallace}},
  \bibinfo{journal}{Physical Review} \textbf{\bibinfo{volume}{71}},
  \bibinfo{pages}{622} (\bibinfo{year}{1947}).

\bibitem[{\citenamefont{Tran et~al.}(2014)\citenamefont{Tran, Soklaski, Liang,
  and Yang}}]{Tran14}
\bibinfo{author}{\bibfnamefont{V.}~\bibnamefont{Tran}},
  \bibinfo{author}{\bibfnamefont{R.}~\bibnamefont{Soklaski}},
  \bibinfo{author}{\bibfnamefont{Y.}~\bibnamefont{Liang}}, \bibnamefont{and}
  \bibinfo{author}{\bibfnamefont{L.}~\bibnamefont{Yang}},
  \bibinfo{journal}{arXiv:1402.4192}  (\bibinfo{year}{2014}).

\bibitem[{\citenamefont{Nair et~al.}(2008)\citenamefont{Nair, Blake,
  Grigorenko, Novoselov, Booth, Stauber, Peres, and Geim.}}]{nair08}
\bibinfo{author}{\bibfnamefont{R.~R.} \bibnamefont{Nair}},
  \bibinfo{author}{\bibfnamefont{P.}~\bibnamefont{Blake}},
  \bibinfo{author}{\bibfnamefont{A.~N.} \bibnamefont{Grigorenko}},
  \bibinfo{author}{\bibfnamefont{K.~S.} \bibnamefont{Novoselov}},
  \bibinfo{author}{\bibfnamefont{T.~J.} \bibnamefont{Booth}},
  \bibinfo{author}{\bibfnamefont{T.}~\bibnamefont{Stauber}},
  \bibinfo{author}{\bibfnamefont{N.~M.~R.} \bibnamefont{Peres}},
  \bibnamefont{and} \bibinfo{author}{\bibfnamefont{A.~K.} \bibnamefont{Geim.}},
  \bibinfo{journal}{Science} \textbf{\bibinfo{volume}{320}},
  \bibinfo{pages}{1308} (\bibinfo{year}{2008}).

\bibitem[{\citenamefont{Mak et~al.}(2008)\citenamefont{Mak, Sfeir, Wu, Lui,
  Misewich, and Heinz}}]{mak08mea}
\bibinfo{author}{\bibfnamefont{K.~F.} \bibnamefont{Mak}},
  \bibinfo{author}{\bibfnamefont{M.~Y.} \bibnamefont{Sfeir}},
  \bibinfo{author}{\bibfnamefont{Y.}~\bibnamefont{Wu}},
  \bibinfo{author}{\bibfnamefont{C.~H.} \bibnamefont{Lui}},
  \bibinfo{author}{\bibfnamefont{J.~A.} \bibnamefont{Misewich}},
  \bibnamefont{and} \bibinfo{author}{\bibfnamefont{T.~F.} \bibnamefont{Heinz}},
  \bibinfo{journal}{Physical Review Letters} \textbf{\bibinfo{volume}{101}},
  \bibinfo{pages}{196405} (\bibinfo{year}{2008}).

\bibitem[{\citenamefont{Qiu et~al.}(2013)\citenamefont{Qiu, Jornada, and
  Louie}}]{qiu13gw}
\bibinfo{author}{\bibfnamefont{D.~Y.} \bibnamefont{Qiu}},
  \bibinfo{author}{\bibfnamefont{F.~H.~d.} \bibnamefont{Jornada}},
  \bibnamefont{and} \bibinfo{author}{\bibfnamefont{S.~G.} \bibnamefont{Louie}},
  \bibinfo{journal}{Physical Review Letters} \textbf{\bibinfo{volume}{111}},
  \bibinfo{pages}{216805} (\bibinfo{year}{2013}).

\bibitem[{\citenamefont{Neto et~al.}(2009)\citenamefont{Neto, Guinea, Peres,
  Novoselov, and Geim.}}]{neto09rmp}
\bibinfo{author}{\bibfnamefont{A.~C.} \bibnamefont{Neto}},
  \bibinfo{author}{\bibfnamefont{F.}~\bibnamefont{Guinea}},
  \bibinfo{author}{\bibfnamefont{N.~M.~R.} \bibnamefont{Peres}},
  \bibinfo{author}{\bibfnamefont{K.~S.} \bibnamefont{Novoselov}},
  \bibnamefont{and} \bibinfo{author}{\bibfnamefont{A.~K.} \bibnamefont{Geim.}},
  \bibinfo{journal}{Review of Modern Physics} \textbf{\bibinfo{volume}{81}},
  \bibinfo{pages}{109} (\bibinfo{year}{2009}).

\bibitem[{\citenamefont{Sarma et~al.}(2011)\citenamefont{Sarma, Adam, Hwang,
  and Rossi}}]{sarma11rmp}
\bibinfo{author}{\bibfnamefont{S.~D.} \bibnamefont{Sarma}},
  \bibinfo{author}{\bibfnamefont{S.}~\bibnamefont{Adam}},
  \bibinfo{author}{\bibfnamefont{E.~H.} \bibnamefont{Hwang}}, \bibnamefont{and}
  \bibinfo{author}{\bibfnamefont{E.}~\bibnamefont{Rossi}},
  \bibinfo{journal}{Review of Modern Physics} \textbf{\bibinfo{volume}{83}},
  \bibinfo{pages}{407} (\bibinfo{year}{2011}).

\bibitem[{\citenamefont{Low et~al.}(2014)\citenamefont{Low, RoldÃ¡n, Han, Xia,
  Avouris, Moreno, and Guinea}}]{low14bpplasmon}
\bibinfo{author}{\bibfnamefont{T.}~\bibnamefont{Low}},
  \bibinfo{author}{\bibfnamefont{R.}~\bibnamefont{RoldÃ¡n}},
  \bibinfo{author}{\bibfnamefont{W.}~\bibnamefont{Han}},
  \bibinfo{author}{\bibfnamefont{F.}~\bibnamefont{Xia}},
  \bibinfo{author}{\bibfnamefont{P.}~\bibnamefont{Avouris}},
  \bibinfo{author}{\bibfnamefont{L.~M.} \bibnamefont{Moreno}},
  \bibnamefont{and} \bibinfo{author}{\bibfnamefont{F.}~\bibnamefont{Guinea}},
  \bibinfo{journal}{arXiv:1404.4035}  (\bibinfo{year}{2014}).

\end{thebibliography}

\end{document}